\documentclass[twocolumn,10pt]{asme2ej}

\usepackage{graphicx} 

\usepackage{pifont}

\usepackage{dingbat}
\usepackage{amsmath}
\usepackage{hyperref}
\usepackage{breqn}
\usepackage{supertabular}
\usepackage{longtable}
\usepackage{float}
\usepackage[font=small,skip=0pt]{caption}

\title{A comprehensive mathematical model of a low-friction servopneumatic actuator}

\author{Rodrigo Trentini\\
        Professor\\
        {\tensfb Daniel Rafael dos Santos}\\
        {\tensfb Yuri Roberto Ferreira}
    \affiliation{
	Undergraduate Research Assistant\\
	Federal Institute of Santa Catarina\\
	Campus Jaraguá do Sul - Rau\\
	Rua dos Imigrantes 405, Jaraguá do Sul\\
    Email: rodrigo.trentini@ifsc.edu.br
    }	
}

\begin{document}

\maketitle    

\begin{abstract}
{\it This paper presents a comprehensive mathematical model of a servopneumatic system, aimed at its consolidation in literature. The work exploits system's friction forces, temperature and pressure evolution, heat transfer, leakage between chambers and environment, equilibrium of cylinder forces, resistance of the pipes, mass flow rate in valve output, varying area of valve orifices and equilibrium of valve forces. Numerical simulations are performed, where system's elastic behaviour due to the air compressibility is thoroughly exploited.
}
\end{abstract}

\vspace{-5mm}

\section{Introduction}

The demand of servopositioning systems driven by compressed air increases day-by-day. Nowadays they have reached a maturity point which enables its control with a fair high accuracy (see, \emph{e.g.}, \cite{Carneiro2012}).

The main criticized point in pneumatics is its energy inefficiency, since the compressors perform two energy transformations (mechanical and thermal) to deliver compressed air to the supply network, as well as its non-linearities  due to friction and air compressibility. However, there are important advantages of servopneumatic systems when compared to electric ones. As examples, one can cite the lower cost, electromagnetic compatibility and the absence of heating.

For using in elastic robots, pneumatic is specially useful due to its intrinsically compliance owing to the air compressibility. Compared to hydraulic actuators, pneumatic ones have indeed less force, but they are cleaner, lighter and less expensive, besides of having the higher power-to-weight ratio over all the off-the-shelf devices.

Nonetheless, differently from electrical machines and despite the interesting works of Carneiro and Almeida \cite{Carneiro2007}, Dunbar \emph{et al.} \cite{Dunbar2001}, Nouri \emph{et al.} \cite{Nouri2000}, Shirazi and Voda \cite{Ravanbod-Shirazi2003} and Richer and Hurmuzlu \cite{Richer2000}, to the best of authors' knowledge there is no unique accepted (and yet accurate) servopneumatic mathematical model. In this sense, this work intends to explore the mathematical modeling and simulation of a servopneumatic actuator towards a wide accepted model.

In order to fulfill this aim, Sec. \ref{chap:general_desc} shows the working principle of cylinders and valves. Section \ref{chap:proposed_servo} shows the proposed solution for a force and positioning control with full mathematical modeling of the system. In Sec. \ref{chap:simulations} the results of the computer simulations are performed, while in Sec. \ref{chap:conclusion} the conclusions of this paper are presented.

\vspace{-5mm}
\section{Working principle}
\label{chap:general_desc}

\begin{figure}[ht]
\centering
\includegraphics[width=0.48\textwidth]{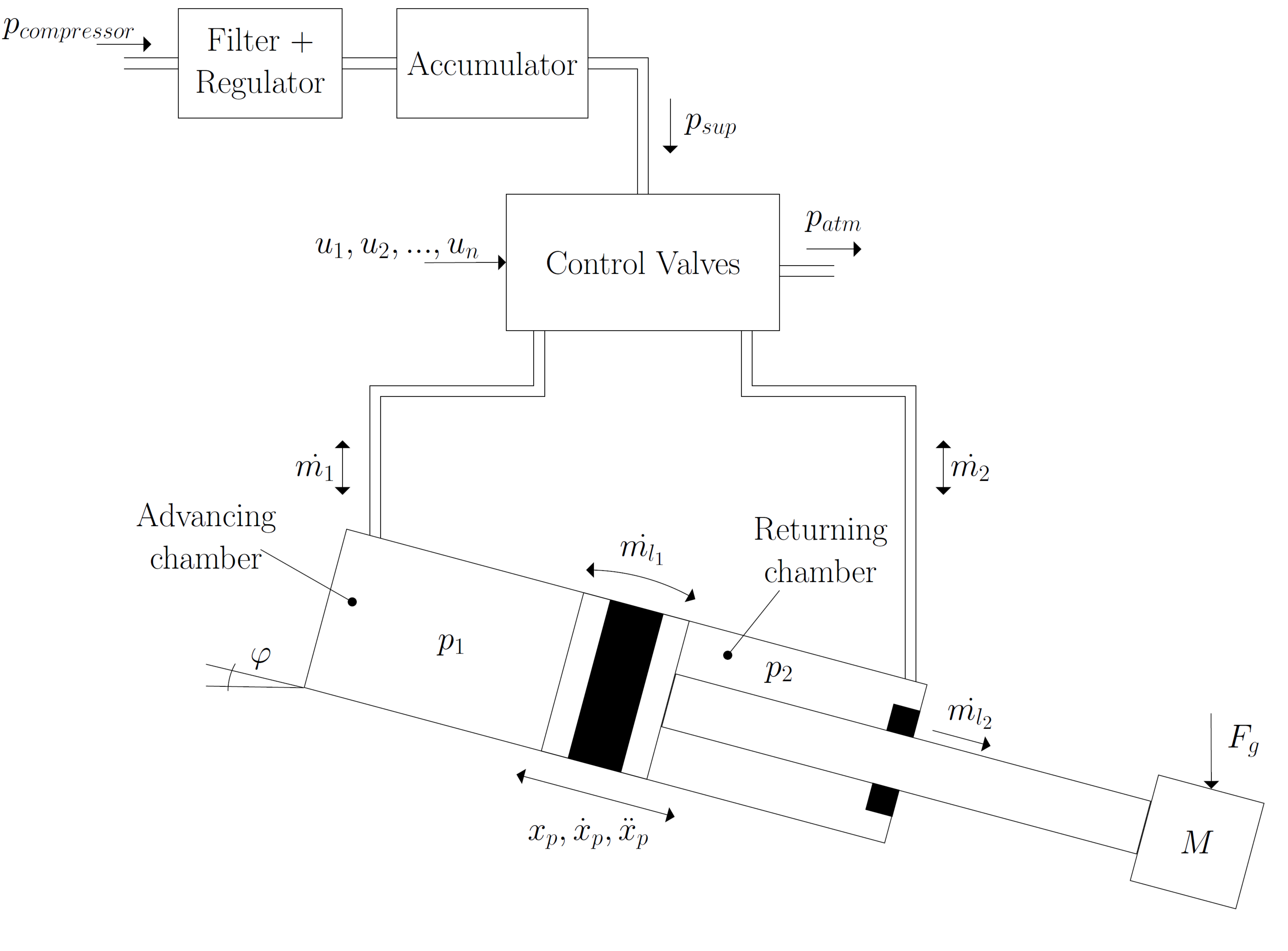} 
\caption{Generic servopneumatic system}
\label{fig:gen_servo}
\end{figure}

A servopneumatic system is described as a positioning system actuated through compressed air, consisted basically of a cylinder, a control unity, sensors and control valve(s). In Fig. \ref{fig:gen_servo} is shown a generic schematic drawing of the described system.

\vspace{-5mm}

\vspace{-7mm}
The system is supplied with a pressure $p_{compressor}$ which feeds the set Filter + Regulator + Accumulator, giving the pressure $p_{sup}$ at the end of this set, which is the supply pressure of the analyzed system. The $n$ control valves are supplied with this pressure and also with the independent control signals $u_1,u_2,\ldots,u_n$. The valves change the mass flow rate $\dot{m}_i$ at their outputs according to the magnitude of the control signals, changing also the pressures $p_1$ and $p_2$ inside the cylinder chambers. These chambers are denominated Advancing and Returning chambers, according Fig. \ref{fig:gen_servo}. It is important to highlight the air leakage between chambers $\dot{m}_{l_1}$ and between the returning chamber and atmosphere $\dot{m}_{l_2}$. The displacement $x_p$ of the cylinder plunger depends on the different between the pressures into the its chambers and also on the areas $A_1$ and $A_2$. When the plunger moves, a certain amount of air can either leave the chamber which has its volume reduced or can be compressed inside the same chamber, depending on the opening of the valve orifice which controls it. On the other hand, a certain amount of air enters the chamber which has its volume increased also depending on the valve which controls it. The leaving air is exhausted te valvo atmosphere ($p_{atm}$) through the same which controls the chamber and has the volume decreased. The load applied to the cylinder rod is submitted to the gravitational force $F_g$. In addition, $\varphi$ represents the inclination angle of the cylinder.

Knowing that friction is the most difficult feature to overcome in such systems, for this work it was chosen a low friction cylinder besides two flow controlled valves, each for one chamber aiming the force control. Despite having a mid air leakage, this drawback is supposed to be easier to deal than the friction effects, which are unpredictable and changeable during time.

Thus, for this implementation a low friction cylinder serie MQM from the manufacturer SMC is chosen. The flow controlled valves are the same as used in Richer and Hurmuzlu \cite{Richer2000}, model PositioneX four-way from Numatics Inc.

\vspace{-5mm}
\section{System's mathematical modeling}
\label{chap:proposed_servo}

The mathematical modeling of this work intends to consider the main features of the servopneumatic system in order to reach an accurate simulation model. It is divided into the dynamics of the cylinder and the valves, besides the effects of the pipes' resistance. Three general considerations must be done: air is considered an ideal gas; the specific heats at constant pressure $c_p$ and constant volume $c_v$ are constant; and there are no losses in the cylinder's orifices.

\vspace{-3mm}
\subsection{Cylinder dynamics}

For the cylinder analysis, it is considered the friction, pressure, temperature, heat transfer and motion dynamics. Also, the leakages between chambers and ambient are considered.

\vspace{-3mm}
\subsubsection{Friction forces}

The friction is known as the major drawback in the control of servopneumatic systems, causing effects like stick-slip and oscillations around the desired position, which can lead the system to instability. For this reason an accurate friction model is important aiming the control of such systems.

An acceptable universal model for friction does not exist yet. Within the many models proposed in literature, one can state that the most used dynamic one is the LuGre, presented by Olsson \emph{et al.} \cite{olsson1998}. This model considers the pre-sliding effect, and it is based on the analysis of the microscopic phenomenon of friction. The model takes into account the displacement of microscopic bristles during pre-sliding region, where the friction dynamics has the form,
\vspace{-3mm}
\begin{equation}
F_f = \sigma_0 z +\sigma_1 \dot{z} +\sigma_2 \dot{x},
\label{eq:ff1}
\end{equation}
\vspace{-5mm}
with,
\vspace{-5mm}
\begin{equation}
\dot{z} = \dot{x} -\frac{\sigma_0 |\dot{x}| z}{g(\dot{x})}.
\label{eq:ff2}
\end{equation}

\vspace{-3mm}
$F_f$ is the friction force, $z$ and $\dot{z}$ are the bristle displacement and velocity, $\dot{x}$ is the relative surface velocity, $\sigma_0$ and $\sigma_1$ are the bristle stiffness and damping, $\sigma_2$ is the viscous friction and $g(\dot{x})$ is a static function that is usually defined as

\vspace{-5mm}
\begin{equation}
g(\dot{x}) = F_c +(F_s -F_c)e^{-\left(\frac{\dot{x}}{v_s} \right)},
\end{equation}

\vspace{-3mm}
\noindent being $F_c$ and $F_s$ the Coulomb and static friction and $v_s$ the Stribeck velocity.

\vspace{-3mm}
\subsubsection{Temperature evolution}

For the pressure dynamics, other considerations must be taken into account: the pressure $p$, temperature $T$ and specific mass $\rho$ are uniform into the chambers, kinetic and gravitational energies are negligible, as well as the viscous work of the fluid. Regard first the state equation for ideal gases $p = \rho R T$. The volumes of advance and returning chambers are considered respectively as,
\vspace{-3mm}
\begin{eqnarray}
V_1 = V_{0_1} +A_1 \left( \frac{L_r}{2} +x \right), \quad V_2 &=& V_{0_2} +A_2 \left( \frac{L_r}{2} -x \right),
\end{eqnarray}

\vspace{-3mm}
\noindent being $V_{0_i}$ the dead volume of the chamber and $L_r$ the total length of the rod.

The First Law of Thermodynamics is applied for the control volume,
\vspace{-7mm}
\begin{equation}
\dot{E} = \dot{W}  + \dot{Q} + \Sigma\dot{m_i}h_i,
\label{eq:energy1}
\end{equation}

\vspace{-3mm}
\noindent being $E$ the energy inside control volume, $W$ the work of external forces and $Q$ the heat transfer with ambient, dot means the time variation of these variables, $\dot{m_i}$ is a mass flow trough control volume and $h_i$ its enthalpy. 

Considering $\dot{E}=\dot{U}=mc_v\dot{T}$, $\dot{W}=-p\dot{V}$, one incoming mass flow $ \dot{m}_{in}$ and one outgoing mass flow $\dot{m}_{out}$, the Eq. \ref{eq:energy1} becomes:

\vspace{-6mm}
\begin{equation}
\begin{split}
m c_v \dot{T} = &-p \dot{V} +\dot{m}_{in} {\left(c_v T_{in} + \frac{p_{in}}{\rho_{in}} + \frac{v_{in}^2}{2}\right)} \\
&- \dot{m}_{out} \left(c_v T_{out} + \frac{p_{out}}{\rho_{out}} +\frac{v_{out}^2}{2}\right) +\dot{Q}.
\end{split}
\end{equation}
\vspace{-3mm}
Now, from the previous hypothesis one can set

\vspace{-5mm}
\begin{equation}
\begin{split}
&c_v T + \frac{p}{\rho} = c_p T = h, \quad \gamma =  \frac{c_p}{c_v}, \quad R = c_p - c_v, \\
&c_p T_{in} + \frac{v^2_{in}}{2} = c_p T_u , \quad
c_p T_{out} + \frac{v^2_{out}}{2} = c_p T, 
\end{split}
\end{equation} 

\vspace{-3mm}
\noindent with $T_u$ the upstream temperature of the fluid considering an infinitely tank so that the flow velocity is zero.

After some algebraic manipulations, one obtain the equation for the temperature evolution inside a varying chamber:

\vspace{-5mm}
\begin{equation}
\dot{T} = \frac{(\gamma -1)T}{P V}\left [-P \dot{V} -RT \dot{m}_{out} + \dot{Q}\right]+ (\gamma T_u -T) \frac{R T}{P V} \dot{m}_{in}.
\label{eq:dT}
\end{equation}

\vspace{-5mm}
\subsubsection{Pressure evolution}

The pressure evolution inside the chambers regards \linebreak $\dot{m} = \dot{m}_{in} -\dot{m}_{out}$. Deriving the state equation for ideal gases:

\vspace{-7mm}
\begin{equation}
\dot{P} = \frac{R T}{V} (\dot{m}_{in} -\dot{m}_{out}) +\frac{P}{T} \dot{T} -\frac{P}{V} \dot{V}.
\label{eq:press1}
\end{equation}

\vspace{-3mm}
Therefore, substituting Eq. \ref{eq:dT} into Eq. \ref{eq:press1} one obtain the final expression of it:

\vspace{-7mm}
\begin{equation}
\dot{P} = -\frac{\gamma P}{V} \dot{V} + \frac{\gamma R T_u}{V} \dot{m}_{in} -\frac{\gamma R T}{V} \dot{m}_{out} + \frac{(\gamma -1)}{V} \dot{Q}.
\label{eq:press_final}
\end{equation}

\vspace{-8mm}
\subsubsection{Heat transfer}

The heat transfer is given directly by Eq. \ref{eq:dQ1}:

\vspace{-7mm}
\begin{equation}
\dot{Q} = -\lambda(P,T) A_q(x) (T -T_a),
\label{eq:dQ1}
\end{equation}

\vspace{-3mm}
\noindent being $\lambda(P,T)$ the coefficient of convective heat transfer between the gas and the wall (heat transfer by radiation and thermal capacity of the chamber walls are neglected), $A_q(x)$ is the surface area of the chamber and $T_a$ is the ambient temperature. $\lambda(P,T)$ is given by,

\vspace{-8mm}
\begin{equation}
\lambda(P,T) = \lambda_0 \left( \frac{P T}{P_0 T_0} \right)^{\frac{1}{2}},
\end{equation}

\vspace{-3mm}
\noindent with $\lambda_0$, $P_0$ and $T_0$ being the heat transfer, pressure and temperature at equilibrium conditions, respectively. As the valve is considered without any internal leakage, $P_0 = P_{atm}$ and $T_0 = T_u$. $A_q(x)$ is given for the chambers as,

\vspace{-5mm}
\begin{equation}
A_q(x) = \left\lbrace 
         \begin{tabular}{l}
         $\pi d_a \left( \frac{L_r}{2} +x \right)$ \text{   for chamber 1}\\
         $\pi d_a \left( \frac{L_r}{2} -x \right)$ \text{   for chamber 2}
		 \end{tabular}          \right.,
\end{equation}

\vspace{-3mm}
\noindent where $d_a$ is the external diameter of the cylinder. It is important to highlight that Eq. \ref{eq:dQ1} considers the heat transfer between the cylinder wall and the environment. The heat transfer between the air inside the chambers and the cylinder wall is not considered, \emph{i.e.} the wall is meant to be at the same temperature as the air inside the respective chamber.

\vspace{-3mm}
\subsubsection{Chambers leakage}

The air leakage between chamber and ambient is modeled as a flow through an orifice following the fluid mechanics analysis, \emph{i.e.} the leakage orifice is considered as a convergent nozzle \cite{Fox2001}.

In this case, two other considerations must be taken into account: there is no friction and the flow is reversible and adiabatic, \emph{i.e.} there is no heat transfer during the flow. The latter can be considered a good approach because the flow is usually fast, and so the heat transfer can be neglected.

Also, the discharging coefficient should be taken into account. Therefore the orifice area given by Eq. \ref{eq:area1} should be multiplied by $c_{d_l}$ due to the \emph{vena contracta} effect.

In such approach there are two different behaviors for the mass flow rate: subsonic and sonic, which are given respectively by:

\vspace{-5mm}
\begin{equation}
\dot{m} = \left\lbrace
          \begin{tabular}{l}
          $p_u c_{d_l} A_l \sqrt{\frac{2 \gamma}{R T_u (\gamma -1)} \left[\left(\frac{p_d}{p_u}\right)^{\frac{2}{\gamma}} -\left(\frac{p_d}{p_u} \right)^{\frac{\gamma+1}{\gamma}}\right]}$ \text{(subsonic)}\\
          $p_u c_{d_l} A_l \sqrt{\frac{\gamma}{R T_u} \left(\frac{2}{\gamma+1} \right)^{\frac{\gamma+1}{\gamma-1}}}$ \hspace{21mm} \text{(sonic).}
          \end{tabular} \right.
\label{eq:flow_rate1}
\end{equation}

\noindent being $p_u$ and $T_u$ the upstream pressure and temperature respectively, $p_d$ the downstream pressure and $A_l$ the leakage area.

For air, the transition between both conditions is given at $p_d/p_u = 0.528$. Above this value the flow is considered chocked (sonic) and under it is consider subsonic.

The up and downstream pressures change according to the magnitude of them: whenever the pressure into one chamber is higher than the other, there exists air leakage between both with the air flowing from the higher to the lower pressure. The same happens for the air leakage from the returning chamber and the environment.

\vspace{-3mm}
\subsubsection{Equilibrium of forces}

Analyzing Fig. \ref{fig:gen_servo} one can obtain the force balance of the system directly through Newton's Second Law:

\vspace{-6mm}
\begin{equation}
M \ddot{x} + F_f + F_{ext} = p_1 A_1 - p_2 A_2 - p_{atm} A_r - F_{hs} - g \sin \varphi,
\label{eq:piston_forces}
\end{equation}

\noindent with $M = M_l +M_p$, where $M_l$ is the load mass, $M_p$ the rod mass, $F_{ext}$ an external force, $A_r$ the rod cross sectional area and $F_{hs}$ the hard stop force due to the cylinder limits, which is modeled as

\vspace{-5mm}
\begin{equation}
F_{hs} = \left\lbrace
         \begin{tabular}{l}
         $0$ \qquad \quad \text{for $x_{min} < x < x_{max}$} \\
         $ p_1 A_1 - p_2 A_2 - p_{atm} A_r - F_f -F_{ext} - g \sin \varphi$ \\ 
        \qquad \qquad \text{for $x_{min} \geq x$ or $x \geq x_{max}$}
         \end{tabular} \right.,
\end{equation}

\noindent in order to ensure zero acceleration when the rod is at one of its limits, \emph{i.e.} $x_{min}$ or $x_{max}$.

\vspace{-3mm}
\subsection{Resistive pipes}

Along the connecting pipes there is a pressure drop due to friction and a time delay on the mass flow due to the speed of the fluid.

The most used equation which describes the behavior of the mass flow rate inside a pipe is given by 1-D Wave, however this is a partial differential equation with the mass flow rate varying according to two variables (time and pipe axis coordinate), it is difficult to handle when one works with complex simulation, since the simulation time increases. 





For this reason, Richer and Hurmuzlu \cite{Richer2000} propose an approximation aiming an ease in the implementation. The proposed model is restricted to small frequencies of the input flow, \emph{i.e.} up to 50 Hz for tubes up to 15 m in length. However, this restriction fits perfectly in servopneumatics applications, since the valve's output frequencies are hardly higher than 15 Hz. It shows that the flow at the outlet of the tube is attenuated in amplitude and delayed by a factor of $L_t/c$, which represents the time required by the input wave to travel through the entire length of the pipe and whose $c$ is the sound speed. This approximation is given by,

\vspace{-5mm}
\begin{equation}
\dot{m}(t) = \left\lbrace
          \begin{tabular}{l}
          $0$ \hspace{29mm} \text{if $t \leq L_t/c$} \\
          $\dot{m}_t(t -\frac{L_t}{c}) e^{-\frac{R_t R T L_t}{2 p c}}$ \hspace{5mm} \text{if $t > L_t/c$}
         \end{tabular} \right.,
\label{eq:mass_flow1}
\end{equation}

\vspace{-3mm}
\noindent where $\dot{m}_t$ is the mass flow rate at pipe's input, $L_t$ is the pipe's length and $R_t$ is the pipe's resistance, which is obtained from the expression for the pressure drop along a tube:

\vspace{-5mm}
\begin{equation}
\Delta p = f \frac{\rho L_t v^2}{2 D} = R_t |v| L_t,
\end{equation}

\vspace{-3mm}
\noindent with $f$ being the friction factor and $D$ the pipe's diameter. After some algebraic manipulations one finds,

\vspace{-5mm}
\begin{equation}
R_t = f \frac{|\dot{m}|}{2 D A_l},
\end{equation}

\vspace{-3mm}
\noindent where $A_l$ is the pipe's cross sectional area.
The friction factor $f$ can be approximated by the Haaland function,

\vspace{-5mm}
\begin{equation}
f = \left\lbrace -1.8 \log_{10} \left[ \frac{6.9}{\text{Re}} + \left( \frac{e_r}{3.7 D} \right)^{1.11} \right]  \right\rbrace^{-2},
\end{equation}

\vspace{-3mm}
\noindent being $e_r$ the roughness of the pipe, Re the Reynold's number and its characteristic length in this case is $D$.

\vspace{-3mm}
\subsection{Servovalve dynamics}

For the valves, the equilibrium of forces, variable area and mass flow rate are analyzed. It is important to highlight that this section is based on Sec. 2.3 of \cite{Richer2000}.

\vspace{-3mm}
\subsubsection{Variable area}

\begin{figure}[H]
\centering
\includegraphics[width=0.33\textwidth]{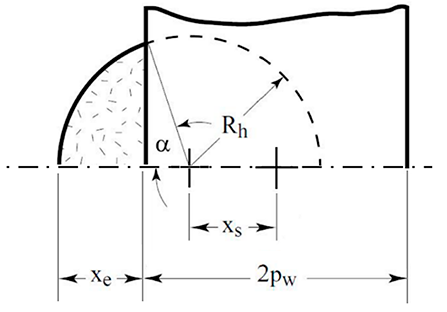} 
\caption{Orifice area versus spool position \cite{Richer2000}}
\label{fig:orificemirror}
\end{figure}

The orifice's variable area of the valves can be represented according to Fig. \ref{fig:orificemirror}.

\vspace{-5mm}

\vspace{-7mm}
The area of the segment of the circle delimited by the edge of the spool can be expressed as,
\begin{equation}
A_e = 2 R_h^2 \arctan \left( \sqrt{\frac{x_e}{2 R_h -x_e}} \right) -(R_h -x_e) \sqrt{x_e (2 R_h -x_e)},
\label{eq:Ae1}
\end{equation}

\noindent where $A_e$ is the effective area for one radial sleeve orifice, $x_e$ is the effective displacement of the valve spool and $R_h$ is the orifice radius. 

The spool width ($2 p_w$) is slightly larger than the radius of the orifice, in order to ensure that the air path is closed even in the presence of small valve misalignment. Thus, the effective displacement of the valve spool $x_e$ will be different from its absolute displacement $x_s$,

\vspace{-7mm}
\begin{equation}
x_e = x_s -(p_w -R_h).
\label{eq:xe}
\end{equation}

\vspace{-3mm}
Substituting Eq. \ref{eq:Ae1} in \ref{eq:xe} the valve effective area for input and exhaust paths becomes respectively:

\vspace{-5mm}
\begin{equation}
A_{e} = \left\lbrace
			 \begin{tabular}{l}
			 $0$ \qquad \qquad \text{if $|x_s| \leq |p_w -R_h|$} \\
			 $2 R_h^2 \tan^{-1} \alpha -(p_w -x_s) \sqrt{R_h^2 -(p_w -x_s)^2}$ \\ \qquad \qquad \hspace{2mm} \text{if $|p_w-R_h|< |x_s|<|p_w+R_h|$} \\
			 $\pi R_h^2$ \qquad \quad \text{if $|x_s| \geq |p_w +R_h|$}
			 \end{tabular} \right.,
\label{eq:area1}			 
\end{equation}

\begin{figure}[H]
Acabei\centering
\includegraphics[width=0.42\textwidth]{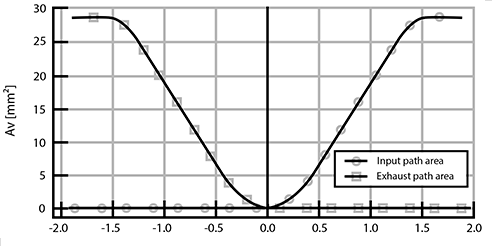} 
\caption{Input and exhaust valve areas \cite{Richer2000}}
\label{fig:graph_orifice}
\end{figure}

\vspace{-3mm}
\noindent being $\alpha = \sqrt{\frac{R_h -p_w +x_s}{R_h +p_w -x_s}}$.
Valve areas for input and exhaust paths versus the spool displacement are presented in Fig. \ref{fig:graph_orifice}.

\vspace{-5mm}

\vspace{-13mm}
\subsubsection{Mass flow rate}

As well as the leakage orifices, the valve orifices can be also considered as convergent nozzles. Thus, Eq. \ref{eq:flow_rate1} is used also in this section.

So, the equations which represent the mass flow rate in the valve output are:

\vspace{-7mm}
\begin{equation}
\dot{m} = \left\lbrace
          \begin{tabular}{l}
          $p_u c_d A_{e_i} \sqrt{\frac{2 \gamma}{R T_u (\gamma -1)} \left[\left(\frac{p}{p_u}\right)^{\frac{2}{\gamma}} -\left(\frac{p}{p_u} \right)^{\frac{\gamma+1}{\gamma}}\right]}$
          \text{(subsonic)}\\
          $p_u c_d A_{e_i} \sqrt{\frac{\gamma}{R T_u} \left(\frac{2}{\gamma+1} \right)^{\frac{\gamma+1}{\gamma-1}}}$ \hspace{22mm} \text{(sonic).}
          \end{tabular}\right.
\label{eq:flow_rate2}
\end{equation}

\vspace{-5mm}
\subsubsection{Equilibrium of forces}

\begin{figure}[H]
\centering
\includegraphics[width=0.3\textwidth]{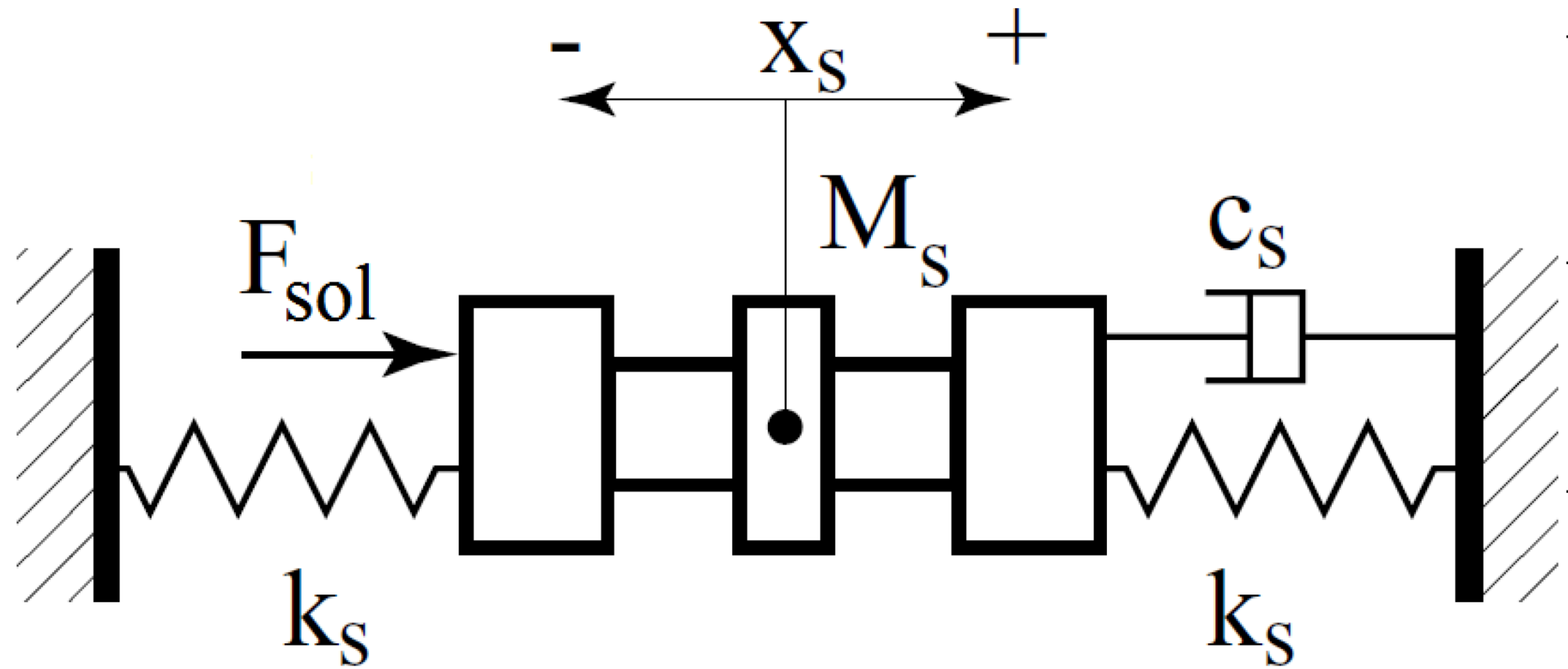} 
\caption{Valve spool dynamic equilibrium \cite{Richer2000}}
\label{fig:valve_schematic}
\end{figure}

In Fig. \ref{fig:valve_schematic} is shown an schematic drawing on the moving parts of the servovalve used in this work.

\vspace{-7mm}

\vspace{-5mm}
In the cited figure, $x_s$ is the displacement of the spool, $k_s$ is the stiffness of the springs, $M_s$ is the mass of the spool, $c_s$ is the viscous friction and $F_{sol}$ is the solenoid force. As can be seen, the friction effect is simplified just to a viscous element. Richer and Hurmuzlu \cite{Richer2000} state that it can be done if it is used a dither electric signal in valve's input, so that the influence of the static friction diminishes.
As well as for the cylinder, the analysis of its dynamics is made directly from Newton's Second Law, being:

\vspace{-8mm}
\begin{equation}
M_s \ddot{x}_s + c_s \dot{x}_s + 2 k_s x_s = K_{sol} i_c,
\label{eq:valve_forces}
\end{equation}

\vspace{-4mm}
\noindent where $K_{sol}$ is the electric constant and $i_c$ is the current on the solenoid.

\vspace{-5mm}
\section{Simulation results}
\label{chap:simulations}


For the implementation of the servopneumatic system through computer simulations, most part of the parameters are chosen according to the cylinder and valve datasheets. Besides, the work of Richer and Hurmuzlu \cite{Richer2000} is used to specify some valve parameters such as $c_s$, $k_s$, $M_s$, $k_{sol}$, $R_h$ and $c_d$. The friction parameters were estimated considering also the cylinder datasheet, as well as the leakage areas.

This work regards the system to be at the sea level, \emph{i.e.} with atmospheric pressure of 1 atm, ambient temperature of 20$^\circ$C and supply pressure of 701325 Pa. Other parameters are omitted for the sake of physical space.


Two different kinds of simulations are defined in order to evaluate the system behavior. These simulations intend to experiment some features of the servopneumatic plant, such as maximum speed and elastic behavior. The first set of simulations intends to analyze the system behavior in high speeds for both directions as well as the influence of load starting (5 kg load). Table \ref{tab:high_speed} shows the parameters for each performed simulation.

\vspace{-7mm}
\begin{table}[H]
	\centering
	\caption{High speed simulations.}
	\scalebox{0.9}{%
		\begin{tabular}{c|c|c|c}
			\hline Simulation	& $i_{c_1}$	[A]	& $i_{c_2}$	[A]	& $M_l$ [kg] 	\\
			\hline (a)-(b)		& 0.5			& -0.5			& 0/5				\\
			\hline (c)-(d)		& -0.5			& 0.5			& 0/5				\\
		\hline
		\end{tabular}}
	\label{tab:high_speed}
\end{table}

\begin{figure}[H]
\centering
\includegraphics[width=0.5\textwidth]{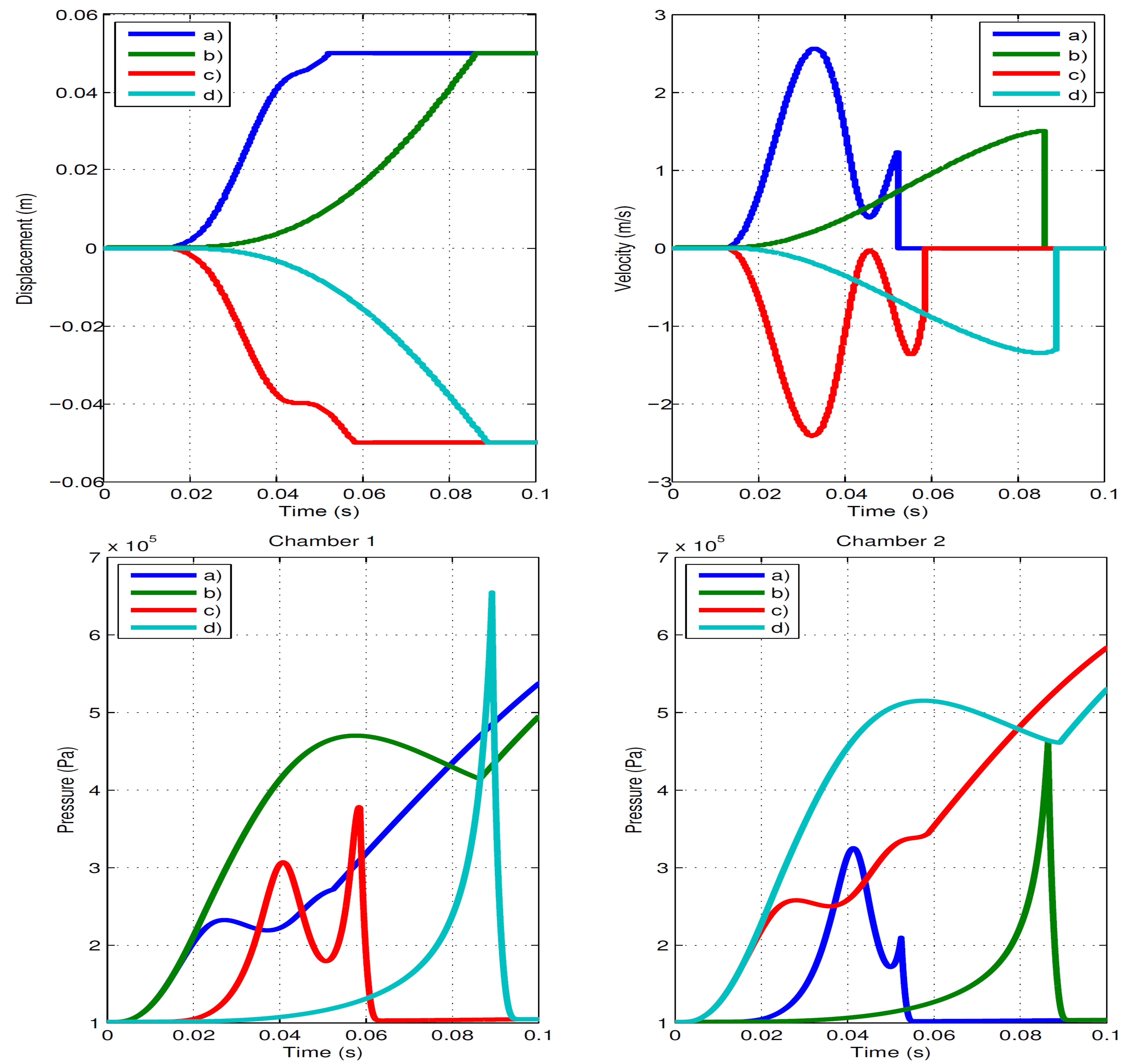} 
\caption{Displacement, velocity and pressures for simulations (a)-(d).}
\label{fig:sim1}
\end{figure}

\vspace{-7mm}
Figure \ref{fig:sim1} shows the displacement, velocity and pressure into both chambers for the cited simulations. It shows that the maximum speed occurs for positive displacement without load and has a magnitude of 2.56 m/s. For negative ones the maximum speed is about 2.41 m/s. For load starting the maximum speed is 1.5 m/s and 1.35 m/s for positive and negative displacements, respectively. Besides, one can see that the system has a dead time of about 20 ms due to friction. It is also possible to see that the pressure begins to increase immediately when the simulations start, \emph{i.e.}, the rod begins to move only when some amount of air is already compressed, increasing the pressure and consequently the force in order to win the static friction.

It is also worth mentioning that for simulations without load, there is an abrupt decrease in the speed near to 40 ms in both directions. It is due to the pressure dynamics, which is slower than the system velocity. In other words, instead of leaving the system, the air inside the chamber which decreases its volume is compressed, leading to an increase on chamber's pressure, reducing rod's speed.

The second set of simulations intend to show the elastic behavior of the plant. The main idea is to hold the system at an equilibrium position without opening the control valves ($i_{c_1} = i_{c_2} = 0$), \emph{i.e.} without letting the air to leave the system (apart from the leakages), and then applying an external force $F_{ext}$. Two different magnitude and duration of the force are simulated: small force with long application time ($t_a$), and great force with short application time (like a hit). Also, the simulations consider the system at equilibrium with different pressures inside its chambers in order to evaluate the influence of compressed air in system's elasticity. Table \ref{tab:elastic} sums up the simulation set.

\vspace{-5mm}

\vspace{-10mm}
\begin{table}[H]
	\centering
	\caption{Elasticity simulations.}
	\scalebox{0.9}{%
	\begin{tabular}{c|c|c|c|c}
			\hline Simulation	& $F_{ext}$	[N]	& $p_1$	[Pa]	& $p_2$ [Pa]	& $t_a$ [s]	\\
			\hline (e)-(f)		& $\pm$20		& $p_a$			& $p_a$			& 0.1		\\
			\hline (g)-(h)		& $\pm$250		& $p_a$			& $p_a$			& 0.01		\\
			\hline (i)-(j)		& $\pm$20		& $p_{sup}$		& 815707		& 0.1		\\
			\hline (k)-(l)		& $\pm$250		& $p_{sup}$		& 815707		& 0.01		\\
		\hline
		\end{tabular}}
	\label{tab:elastic}
\end{table}

\begin{figure}[ht]
\centering
\includegraphics[width=0.5\textwidth]{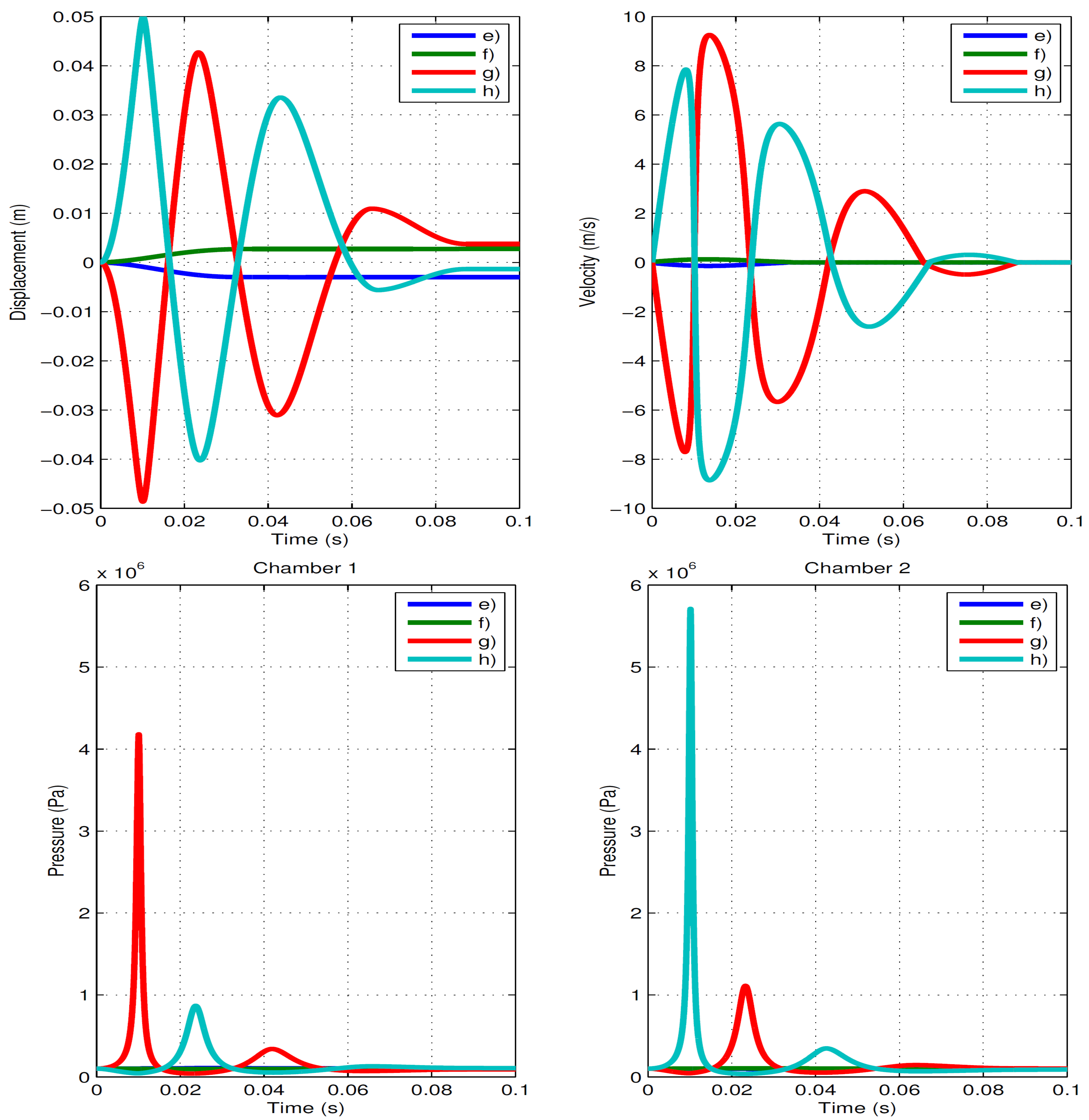} 
\caption{Displacement, velocity and pressures for simulations (e)-(h).}
\label{fig:sim2}
\end{figure}

\vspace{-6mm}
Figure \ref{fig:sim2} shows simulations (e)-(h). The first shows the position and velocity of cylinder's rod, while the last shows the pressure evolution in both chambers. Notice that the rod stops moving around 30 ms for simulations (e) and (f) even with the force $F_{ext}$ being applied. It occurs because the applied force (20 N) is not enough to win the internal force produced by the air compression inside the chamber, and this is the reason for not having oscillations at rod's position. For all simulations, the rod does not return to the former equilibrium position at $x = 0$ m. On the hand, the system reaches a new equilibrium point due to two factors: the friction and mainly the pressure dynamics. The oscillations obtained from the ``hit'' simulations prove the elastic feature due to air compressibility. Moreover, the pressure into the returning chamber increases more than in chamber 1 because of the difference in the areas, reaching almost 6 MPa for the ``hit'' simulation in negative direction.


\begin{figure}[ht]
\centering
\includegraphics[width=0.5\textwidth]{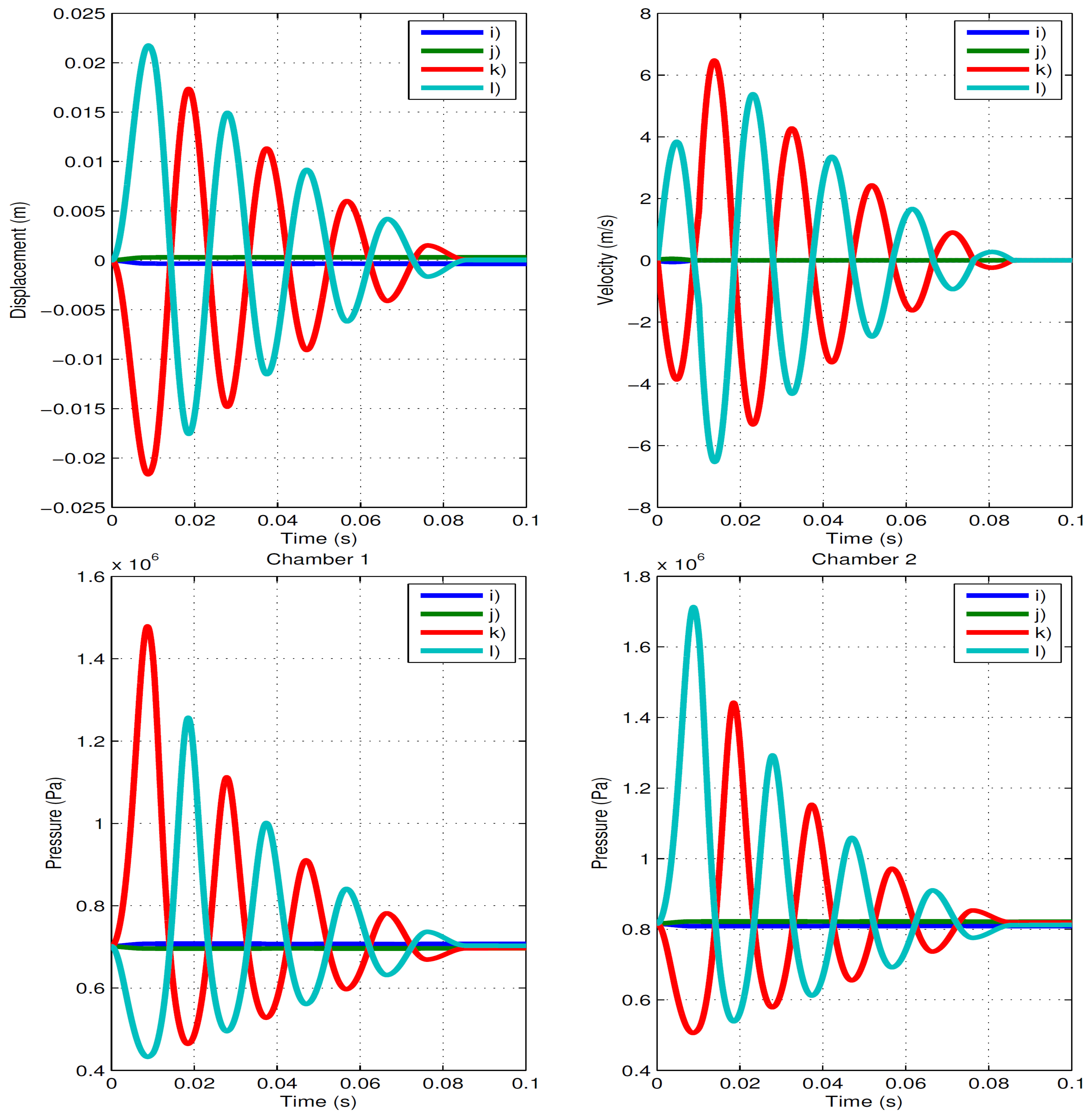} 
\caption{Displacement, velocity and pressures for simulations (i)-(l).}
\label{fig:sim3}
\end{figure}

In Fig. \ref{fig:sim3} one can observe the simulation results with pressurized system (items (i)-(l)). All the points cited for simulations (e)-(h) can be stated again, but with some other new considerations. For instance, the amplitude of the motion decreased in a factor of around 2.3, and the final equilibrium position is much closer to $x=0$ m than the previous simulations. At last, the motion as well as the pressure in both chambers are much more oscillatory for the ``hit'' simulations.

\vspace{-4mm}

It is justified by the fact that for these simulations the chambers were previously pressurized, which makes it harder to move when an external force is applied. Also, the damping decreases due to the same reason, \emph{i.e.} the system response to any disturbance is faster because the pressure is higher, causing more oscillations at rod's position.

\vspace{-4mm}	
\section{Conclusions}
\label{chap:conclusion}

This work has shown a comprehensive mathematical modeling and simulation of a low friction servopneumatic system. Several features of the system were considered, such as the friction forces, temperature and pressure evolution, heat transfer, leakage between chambers and environment, equilibrium of cylinder's forces, the resistance of the pipes, the mass flow rate at the valve output, the varying area of the valve orifices and the equilibrium of valve's forces. Numerical simulations have shown that the proposed system is able to reach a maximum speed of 2.56 m/s for the applied supply pressure $p_{sup}$, besides of an elastic behavior due to the air compressibility.

\vspace{-4mm}
\section{Acknowledgement}

The authors acknowledge the support of the Santa Catarina State Research Support Foundation (FAPESC), process IFS2018121000005.

\vspace{-4mm}	
\bibliographystyle{asmems4}

\bibliography{bib}



\end{document}